\documentclass[12pt,english,aps,superscriptaddress,preprintnumbers]{revtex4-1}
\usepackage[T1]{fontenc}
\usepackage[latin9]{inputenc}
\usepackage{amsmath}
\usepackage{amssymb}
\usepackage{esint}

\makeatletter

\usepackage{mathrsfs}

\@ifundefined{textcolor}{}{%
 \definecolor{BLACK}{gray}{0}
 \definecolor{WHITE}{gray}{1}
 \definecolor{RED}{rgb}{1,0,0}
 \definecolor{GREEN}{rgb}{0,1,0}
 \definecolor{BLUE}{rgb}{0,0,1}
 \definecolor{CYAN}{cmyk}{1,0,0,0}
 \definecolor{MAGENTA}{cmyk}{0,1,0,0}
 \definecolor{YELLOW}{cmyk}{0,0,1,0}
}

\usepackage{babel}

\makeatother

\usepackage{babel}
\begin{document}
\title{Calculation of ground state energy of Lithium and Beryllium based
on variational method}
\author{Mei-Qin Deng}
\affiliation{Department of Physics, College of Intelligent Systems Science and
Engineering, Hubei Minzu University, Enshi, Hubei 445000, China}
\author{Ren-Hong Fang}
\email{fangrh@sdu.edu.cn}

\affiliation{Department of Physics, College of Intelligent Systems Science and
Engineering, Hubei Minzu University, Enshi, Hubei 445000, China}
\begin{abstract}
With the consideration of identity principle and atomic shell structure,
we calculated the ground state energy of Lithium atom and Beryllium
atom based on variational method, which accords quite well with the
experimental results.
\end{abstract}
\maketitle

\section{Introduction}

In quantum mechanics, it is hard to solve many-body systems analytically.
For many-electron atoms, variational method is often adopted to calculate
energy levels of atoms approximately \citep{16Mao:2006ois,1Deng:2013ois,10Quan:2014ois,2Yang:2014ois,15Zhang:2015ois,3Lu:2019ois,9Yin:2021ois,14Pi:2023ois}.
Calculation of ground state energy of Helium atom can be found in
traditional textbooks \citep{Qian:2006ac,Zhou:2022ac}. For Lithium
and Beryllium atoms, the calculation of ground state energy is more
complex than Helium atom \citep{8Wang:1993ois,7Kong:1995ois,17Saleh:1998ois,5Huang:1999ois,12Li:2002ois,4Wu:2017ois,6Dong:2018ois}.
There are also some papers calculating the energy levels of atoms
with atomic number $Z>4$, where the calculation procedures are extremely
complicated \citep{13Li:2006ois,11Huang:2007ois}. In this article,
we will introduce a simplified variational method, with the consideration
of identity principle and atomic shell structure, to calculate the
ground state energy of Lithium and Beryllium atoms.

Now we briefly introduce the variational method in quantum mechanics.
The Hamiltonian operator of a system is denoted as $\hat{H}$, whose
eigenvalues and corresponding eigenfunctions are listed in the following,
\[
E_{0}<E_{1}<E_{2}<\cdots<E_{n}<\cdots
\]
\[
\psi_{0},\ \psi_{1},\ \psi_{2},\cdots,\ \psi_{n},\cdots
\]
According to the completeness theorem of the eigenfunctions of an
Hermitian operator, any normalised wavefunction $\Phi$ can be expanded
by $\{\psi_{n}\}$ as follows,
\begin{equation}
\Phi=\sum_{n}C_{n}\psi_{n}\label{eq:q1}
\end{equation}
The average value of $\hat{H}$ in $\Phi$ state is 
\begin{equation}
\overline{\hat{H}}=\int dV\Phi^{*}\hat{H}\Phi=\sum_{n=0}^{\infty}E_{n}|C_{n}|^{2}\geqslant E_{0}\sum_{n=0}^{\infty}|C_{n}|^{2}=E_{0}\label{eq:q2}
\end{equation}
i.e., the average value of $\hat{H}$ in any normalised state is bigger
than the ground state energy $E_{0}$. If we introduce a parameter
$\lambda$ in the normalised wavefunction $\Phi$ (called trial wavefunction),
then the average value of $\hat{H}$ in $\Phi$ state can be written
as $\overline{\hat{H}}(\lambda)$, which depends on the parameter
$\lambda$. Solving $d\overline{\hat{H}}(\lambda)/d\lambda=0$, we
can find the extreme point $\lambda_{0}$, which gives local minimum
$\overline{\hat{H}}(\lambda_{0})$. Then $\overline{\hat{H}}(\lambda_{0})$
can be chosen as the approximate value of the ground state energy
$E_{0}$. 

In this article, we will use variational method to calculate the ground
state energy of Lithium and Beryllium atoms. For a neutral atom with
atomic number $Z$, it has $Z$ electrons outside of the nucleus.
These electrons are identical, which leads to the antisymmetric wavefunction
of electrons. In this article, the electrons outside of the nucleus
are assumed to move independently in the mean Coulomb field. For $N$
electrons in $N$ different states, the antisymmetric wavefunction
is
\begin{equation}
\Phi(1,2,\cdots,N)=\frac{1}{\sqrt{N!}}\left|\begin{array}{cccc}
\phi_{i}(1) & \phi_{i}(2) & \cdots & \phi_{i}(N)\\
\phi_{j}(1) & \phi_{j}(2) & \cdots & \phi_{j}(N)\\
\vdots & \vdots & \vdots & \vdots\\
\phi_{k}(1) & \phi_{k}(2) & \cdots & \phi_{k}(N)
\end{array}\right|=\frac{1}{\sqrt{N!}}\sum_{P}\delta_{P}P\left[\phi_{i}(1)\phi_{j}(2)\cdots\phi_{k}(N)\right]\label{eq:q3}
\end{equation}
where $P$ represents a kind of permutation, $\delta_{P}=1$ for even
permutation and $\delta_{P}=-1$ for odd permutation. Besides the
antisymmetry of wavefunction, the shell structure of atoms is also
considered in this article.

The rest of this article is organized as follows. In Sec. II, the
ground state energy of Lithium atom is calculated. In Sec. III, the
ground state energy of Beryllium atom is calculated. We summarize
this article in Sec. IV. Some calculation details are listed in the
appendix.

\section{Ground state energy of Lithium}

\label{sec:Li}Lithium atom has three electrons outside of the nucleus.
The heavy nucleus can be set in the coordinate origin. The Hamiltonian
operator of the electrons is 
\begin{equation}
\hat{H}=\sum_{i=1}^{Z}\left(-\frac{\hbar^{2}}{2m}\nabla_{i}^{2}\right)+\sum_{i=1}^{Z}\left(-\frac{Ze^{2}}{4\pi\varepsilon_{0}r_{i}}\right)+\sum_{i<j}\frac{e^{2}}{4\pi\varepsilon_{0}r_{ij}}\label{eq:q4}
\end{equation}
where $Z=3$ for Lithium atom. For the ground state of Lithium atom,
two electrons are in $1s$ state (with spin up and down respectively),
and the third electron is in $2s$ state (Suppose spin up). The wavefunctions
of these three states are denoted as $\phi_{a},\phi_{b},\phi_{c}$:
\begin{equation}
\phi_{a}=\psi_{100}\chi_{+},\ \ \ \phi_{b}=\psi_{100}\chi_{-},\ \ \ \phi_{c}=\psi_{200}\chi_{+}\label{eq:q5}
\end{equation}
The explicit forms of $\psi_{100},\psi_{200}$ and $\chi_{+},\chi_{-}$
are listed as follows,
\begin{equation}
\psi_{100}(r)=\sqrt{\frac{\lambda^{3}}{\pi a_{B}^{3}}}e^{-\lambda r/a_{B}},\ \ \ \psi_{200}(r)=\sqrt{\frac{\lambda^{3}}{\pi a_{B}^{\prime3}}}\left(1-\frac{\lambda r}{a_{B}^{\prime}}\right)e^{-\lambda r/a_{B}^{\prime}}\label{eq:1}
\end{equation}
\begin{equation}
\chi_{+}=\left(\begin{array}{c}
1\\
0
\end{array}\right),\ \ \ \chi_{-}=\left(\begin{array}{c}
0\\
1
\end{array}\right)\label{eq:2}
\end{equation}
where the parameter $\lambda$ is the effective charge felt by the
three electrons due to screening effect, $a_{B}$ is Bohr radius,
and $a_{B}^{\prime}=2a_{B}$.

To calculate the ground state energy of Lithium atom based on variational
method, we must choose a trial wavefunction. The three electrons outside
of the nucleus are identical fermions, so the trial wavefunction should
satisfy exchange antisymmetry. The trial wavefunction (normalised)
can be chosen as 
\begin{eqnarray}
\Psi & = & \frac{1}{\sqrt{3!}}\sum_{P}\delta_{P}P\left[\phi_{a}(1)\phi_{b}(2)\phi_{c}(3)\right]\nonumber \\
 & = & \frac{1}{\sqrt{3!}}\bigg[\phi_{a}(1)\phi_{b}(2)\phi_{c}(3)+\phi_{a}(2)\phi_{b}(3)\phi_{c}(1)+\phi_{a}(3)\phi_{b}(1)\phi_{c}(2)\nonumber \\
 &  & -\phi_{a}(2)\phi_{b}(1)\phi_{c}(3)-\phi_{a}(3)\phi_{b}(2)\phi_{c}(1)-\phi_{a}(1)\phi_{b}(3)\phi_{c}(2)\bigg]\label{eq:q6}
\end{eqnarray}

The average value of $\hat{H}$ in $\Psi$ state is 
\begin{equation}
\overline{\hat{H}}(\lambda)=\int\Psi^{\dagger}\hat{H}\Psi dV_{1}dV_{2}dV_{3}\label{eq:q7}
\end{equation}
Firstly we calculate $\Psi^{\dagger}\Psi$. We adopt following symbol
to simplify calculation:
\begin{equation}
(ijk)\equiv\phi_{a}(i)\phi_{b}(j)\phi_{c}(k)\label{eq:q8}
\end{equation}
then the trial wavefunction $\Psi$ can be rewritten as 
\begin{equation}
\Psi=\frac{1}{\sqrt{3!}}[(123)+(231)+(312)-(213)-(321)-(132)]\label{eq:q9}
\end{equation}
The result of $\Psi^{\dagger}\Psi$ is
\begin{eqnarray}
\Psi^{\dagger}\Psi\times3! & = & (123)^{2}-(123)(321)+(231)^{2}-(231)(132)+(312)^{2}-(312)(213)\nonumber \\
 &  & +(213)^{2}-(213)(312)+(321)^{2}-(321)(123)+(132)^{2}-(132)(231)\nonumber \\
 & = & 2\times[(123)^{2}+(231)^{2}+(312)^{2}]\nonumber \\
 &  & -2\times[(123)(321)+(231)(132)+(312)(213)]\label{eq:w1}
\end{eqnarray}
where $(123)(321)$ means $(123)^{\dagger}(321)$, and $(123)^{2}$
means $(123)^{\dagger}(123)$, etc. 

Making use of the virial theorem in quantum mechanics, the average
value of the kinetic energy operators and the potential energy between
nucleus and electron of $\hat{H}$ in $\Psi$ state can be easily
obtained:
\begin{equation}
\overline{\sum_{i=1}^{Z}\left(-\frac{\hbar^{2}}{2m}\nabla_{i}^{2}\right)}=Z\times\overline{\left(-\frac{\hbar^{2}}{2m}\nabla_{1}^{2}\right)}=\frac{3}{8}Z\lambda^{2}\alpha^{2}mc^{2}\label{eq:w2}
\end{equation}
\begin{equation}
\overline{\sum_{i=1}^{Z}\left(-\frac{Ze^{2}}{4\pi\varepsilon_{0}r_{i}}\right)}=Z\times\overline{\left(-\frac{Ze^{2}}{4\pi\varepsilon_{0}r_{1}}\right)}=-\frac{3}{4}Z^{2}\lambda\alpha^{2}mc^{2}\label{eq:w3}
\end{equation}
where $\alpha$ is fine structure constant and $mc^{2}$ is the rest
energy of an electron. In $\Psi$ state, the average value of the
potential energy among the three electrons of $\hat{H}$ is 
\begin{eqnarray}
\overline{\sum_{i<j}\frac{e^{2}}{4\pi\varepsilon_{0}r_{ij}}} & = & 3\times\overline{\left(\frac{e^{2}}{4\pi\varepsilon_{0}r_{12}}\right)}\nonumber \\
 & = & \alpha\hbar c\int dV_{1}dV_{2}\frac{1}{r_{12}}\bigg[\psi_{100}(r_{1})^{2}\psi_{100}(r_{2})^{2}+2\times\psi_{100}(r_{1})^{2}\psi_{200}(r_{2})^{2}\nonumber \\
 &  & -\psi_{100}(r_{1})\psi_{100}(r_{2})\psi_{200}(r_{1})\psi_{200}(r_{2})\bigg]\nonumber \\
 & = & \frac{5965}{5832}\times\lambda\alpha^{2}mc^{2}\label{eq:w4}
\end{eqnarray}
where the three six-dimensional integrals have been calculated in
appendix A. Here we list the three integrals:
\begin{equation}
\int dV_{1}dV_{2}\frac{1}{r_{12}}\psi_{100}(r_{1})^{2}\psi_{100}(r_{2})^{2}=\frac{5}{8}\times\frac{\lambda}{a_{B}}\label{eq:w6}
\end{equation}
\begin{equation}
\int dV_{1}dV_{2}\frac{1}{r_{12}}\psi_{100}(r_{1})^{2}\psi_{200}(r_{2})^{2}=\frac{17}{81}\times\frac{\lambda}{a_{B}}\label{eq:w8}
\end{equation}
\begin{equation}
\int dV_{1}dV_{2}\frac{1}{r_{12}}\psi_{100}(r_{1})\psi_{100}(r_{2})\psi_{200}(r_{1})\psi_{200}(r_{2})=\frac{16}{27^{2}}\times\frac{\lambda}{a_{B}}\label{eq:w5}
\end{equation}
The first two integrals contribute to Coulomb energy, and the third
term comes from exchange effect due to the antisymmetric wavefunction. 

Finally we obtain the average value of $\hat{H}$ in $\Psi$ state
as follows,
\begin{equation}
\overline{\hat{H}}(\lambda)=\alpha^{2}mc^{2}\times\left(\frac{9}{8}\lambda^{2}-\frac{33\,401}{5832}\lambda\right)\label{eq:w9}
\end{equation}
Setting $d\overline{\hat{H}}(\lambda)/d\lambda=0$, one get 
\begin{equation}
\lambda_{0}=\frac{33\,401}{13\,122}\approx2.5454\label{eq:e1}
\end{equation}
which gives the local minimum,
\begin{equation}
\overline{\hat{H}}(\lambda_{0})=-\frac{1\,115\,626\,801}{153\,055\,008}\times\alpha^{2}mc^{2}\approx-198.345\,\mathrm{eV}\label{eq:e2}
\end{equation}
This is the ground state energy of Lithium atom obtained through variational
method, which differs from the experimental value $-203.48\,\mathrm{eV}$
by relative error $2.5\%$. 

\section{Ground state energy of Beryllium}

\label{sec:Be}Beryllium atom has four electrons outside of the nucleus.
For the ground state of a Beryllium atom, two electrons are in $1s$
state (with spin up and down respectively), and the other two electrons
are in $2s$ state (with spin up and down respectively). The wavefunctions
of these four states are denoted as $\phi_{a},\phi_{b},\phi_{c},\phi_{d}$:
\begin{equation}
\phi_{a}=\psi_{100}\chi_{+},\ \ \ \phi_{b}=\psi_{100}\chi_{-},\ \ \ \phi_{c}=\psi_{200}\chi_{+},,\ \ \ \phi_{d}=\psi_{200}\chi_{-}\label{eq:e3}
\end{equation}
The explicit forms of $\psi_{100},\psi_{200}$ and $\chi_{+},\chi_{-}$
are the same as Eqs. (\ref{eq:1})-(\ref{eq:2}). Now the parameter
$\lambda$ is the effective charge felt by the four electrons due
to screening effect.

The trial antisymmetric wavefunction (normalised) of the four electrons
can be chosen as 
\begin{eqnarray}
\Psi & = & \frac{1}{\sqrt{4!}}\sum_{P}\delta_{P}P\left[\phi_{a}(1)\phi_{b}(2)\phi_{c}(3)\phi_{d}(4)\right]\nonumber \\
 & = & \frac{1}{\sqrt{4!}}\bigg[(1234+2314+3124-2134-3214-1324)\nonumber \\
 &  & -(1243+2341+3142-2143-3241-1342)\nonumber \\
 &  & +(1423+2431+3412-2413-3421-1432)\nonumber \\
 &  & -(4123+4231+4312-4213-4321-4132)\bigg]\label{eq:e4}
\end{eqnarray}
The average value of $\hat{H}$ in $\Psi$ state is 
\begin{equation}
\overline{\hat{H}}(\lambda)=\int\Psi^{\dagger}\hat{H}\Psi dV_{1}dV_{2}dV_{3}dV_{4}\label{eq:e5}
\end{equation}
The result of $\Psi^{\dagger}\Psi$ is
\begin{eqnarray}
 &  & \Psi^{\dagger}\Psi\times4!\nonumber \\
 & = & \ \ (1234)^{2}+1234(-3214-1432+3412)+(2314)^{2}+2314(-1324-2413+1423)\nonumber \\
 &  & +(3124)^{2}+3124(-2134-3421+2431)+(2134)^{2}+2134(-3124-2431+3421)\nonumber \\
 &  & +(3214)^{2}+3214(-1234-3412+1432)+(1324)^{2}+1324(-2314-1423+2413)\nonumber \\
 &  & +(\mathrm{other\ 72\ terms})\label{eq:e6}
\end{eqnarray}
There are totally $24^{2}=576$ terms in $\Psi^{\dagger}\Psi$, however
with only $96$ terms nonzero. The complete form of $\Psi^{\dagger}\Psi$
is listed in Appendix B.

The average value of the kinetic energy operators of $\hat{H}$ in
$\Psi$ state is 
\begin{equation}
\overline{\sum_{i=1}^{Z}\left(-\frac{\hbar^{2}}{2m}\nabla_{i}^{2}\right)}=Z\times\overline{\left(-\frac{\hbar^{2}}{2m}\nabla_{1}^{2}\right)}=\frac{5}{16}Z\lambda^{2}\alpha^{2}mc^{2}\label{eq:e7}
\end{equation}
with $Z=4$. In $\Psi$ state, the average value of the potential
energy between nucleus and the three electrons of $\hat{H}$ is 
\begin{equation}
\overline{\sum_{i=1}^{Z}\left(-\frac{Ze^{2}}{4\pi\varepsilon_{0}r_{i}}\right)}=Z\times\overline{\left(-\frac{Ze^{2}}{4\pi\varepsilon_{0}r_{1}}\right)}=-\frac{5}{8}Z^{2}\lambda\alpha^{2}mc^{2}\label{eq:e8}
\end{equation}
The average value of the potential energy among the four electrons
of $\hat{H}$ in $\Psi$ state is 
\begin{eqnarray}
\overline{\sum_{i<j}\frac{e^{2}}{4\pi\varepsilon_{0}r_{ij}}} & = & 3\times\overline{\left(\frac{e^{2}}{4\pi\varepsilon_{0}r_{12}}\right)}\nonumber \\
 & = & 6\alpha\hbar c\times\frac{1}{4!}\int dV_{1}dV_{2}\frac{1}{r_{12}}\bigg[4\times\psi_{100}(r_{1})^{2}\psi_{100}(r_{2})^{2}+4\times\psi_{200}(r_{1})^{2}\psi_{200}(r_{2})^{2}\nonumber \\
 &  & +16\times\psi_{100}(r_{1})^{2}\psi_{200}(r_{2})^{2}-8\times\psi_{100}(r_{1})\psi_{200}(r_{1})\psi_{100}(r_{2})\psi_{200}(r_{2})\bigg]\nonumber \\
 & = & \frac{586\,373}{373\,248}\times\lambda\alpha^{2}mc^{2}\label{eq:e9}
\end{eqnarray}
In appendix A, we have calculated following integral:
\begin{equation}
\int dV_{1}dV_{2}\frac{1}{r_{12}}\psi_{200}(r_{1})^{2}\psi_{200}(r_{2})^{2}=\frac{77}{512}\times\frac{\lambda}{a_{B}}\label{eq:r1}
\end{equation}

Finally we obtain the average value of $\hat{H}$ in $\Psi$ state
as follows,
\begin{equation}
\overline{\hat{H}}(\lambda)=\alpha^{2}mc^{2}\times\left(\frac{5}{4}\lambda^{2}-\frac{3\,146\,107}{373\,248}\lambda\right)\label{eq:r2}
\end{equation}
Setting $d\overline{\hat{H}}(\lambda)/d\lambda=0$, one get 
\begin{equation}
\lambda_{0}=\frac{3\,146\,107}{933\,120}\approx3.3716\label{eq:r3}
\end{equation}
which gives the local minimum: 
\begin{equation}
\overline{\hat{H}}(\lambda_{0})=-\frac{9\,897\,989\,255\,449}{696\,570\,347\,520}\times\alpha^{2}mc^{2}\approx-386.663\,\mathrm{eV}\label{eq:r4}
\end{equation}
This is the ground state energy of Beryllium atom obtained through
variational method, which differs from the experimental value $-399.14\,\mathrm{eV}$
by relative error $3.1\%$. 

\section{Summary}

\label{sec:Summary}A neutral atom with atomic number $Z$ has $Z$
electrons outside of the nuclear. These electrons are identical, which
leads to the antisymmetry of wavefunction according to identity principle.
With the consideration of antisymmetric wavefunction and atomic shell
structure, theoretical results of the ground state energy of Lithium
atom ($Z=3$) and Beryllium atom ($Z=4$) are obtained through variational
method, which agree with the experimental results quite well. For
atoms with $Z>4$, the calculation of ground state energy is more
complicated than Lithium and Beryllium atoms, and the help from computer
programs may be necessary.

\section{\textup{Acknowledgments}}

This work is supported in part by the Science and Technology Research
Project of Education Department of Hubei Province of China (Grant
No. D20221901), the National Natural Science Foundation of China (Grants
No. 12265013 and No. 12473074), and the Program for Innovative Youth
Research Team in University of Hubei Province of China (Grant No.
T201712).

\appendix

\section{Four six-dimensional integrals }

\label{sec:4-integrals}In this section, we will calculate following
four six-dimensional integrals:
\begin{eqnarray}
\mathrm{I}_{1} & = & \int dV_{1}dV_{2}\frac{1}{r_{12}}\psi_{100}(r_{1})^{2}\psi_{100}(r_{2})^{2}\nonumber \\
\mathrm{I}_{2} & = & \int dV_{1}dV_{2}\frac{1}{r_{12}}\psi_{100}(r_{1})^{2}\psi_{200}(r_{2})^{2}\nonumber \\
\mathrm{I}_{3} & = & \int dV_{1}dV_{2}\frac{1}{r_{12}}\psi_{100}(r_{1})\psi_{100}(r_{2})\psi_{200}(r_{1})\psi_{200}(r_{2})\nonumber \\
\mathrm{I}_{4} & = & \int dV_{1}dV_{2}\frac{1}{r_{12}}\psi_{200}(r_{1})^{2}\psi_{200}(r_{2})^{2}\label{eq:A1}
\end{eqnarray}
where the explicit forms of $\psi_{100}(r_{1}),\psi_{100}(r_{2})$
are
\begin{equation}
\psi_{100}(r)=\sqrt{\frac{\lambda^{3}}{\pi a_{B}^{3}}}e^{-\lambda r/a_{B}},\ \ \ \psi_{200}(r)=\sqrt{\frac{\lambda^{3}}{\pi a_{B}^{\prime3}}}\left(1-\frac{\lambda r}{a_{B}^{\prime}}\right)e^{-\lambda r/a_{B}^{\prime}}\label{eq:s1}
\end{equation}

For these six-dimensional integrals, we integrate $\boldsymbol{r}_{1}$
firstly, then integrate $\boldsymbol{r}_{2}$. When integrating $\boldsymbol{r}_{1}$,
we set $\boldsymbol{r}_{2}$ on the positive direction of $z$-axis.
The $\frac{1}{r_{12}}$ term can be expanded as follows \citep{Liang:1995ac}:
\begin{equation}
\frac{1}{r_{12}}=\frac{1}{|\boldsymbol{r}_{1}-\boldsymbol{r}_{2}|}=\left\{ \begin{array}{cc}
\frac{1}{r_{2}}\sum_{l=0}^{\infty}\left(\frac{r_{1}}{r_{2}}\right)^{l}P_{l}(\cos\theta), & r_{1}<r_{2}\\
\frac{1}{r_{1}}\sum_{l=0}^{\infty}\left(\frac{r_{2}}{r_{1}}\right)^{l}P_{l}(\cos\theta), & r_{1}>r_{2}
\end{array}\right.\label{eq:3s2}
\end{equation}
where $\theta$ is the included angle between $\boldsymbol{r}_{1}$
and $\boldsymbol{r}_{2}$. Firstly, we calculate following integral:
\[
\int dV_{1}\frac{1}{r_{12}}f(r_{1})
\]
where $f(r_{1})$ is the function of $r_{1}=|\boldsymbol{r}_{1}|$.
Making use of Eq. (\ref{eq:3s2}), we can obtain
\begin{eqnarray}
\int dV_{1}\frac{1}{r_{12}}f(r_{1}) & = & \int_{r_{1}<r_{2}}dV_{1}\frac{1}{r_{12}}f(r_{1})+\int_{r_{1}>r_{2}}dV_{1}\frac{1}{r_{12}}f(r_{1})\nonumber \\
 & = & \int_{0}^{r_{2}}dr_{1}r_{1}^{2}\frac{1}{r_{2}}\sum_{l=0}^{\infty}\left(\frac{r_{1}}{r_{2}}\right)^{l}f(r_{1})\int P_{l}(\cos\theta)d\Omega\nonumber \\
 &  & +\int_{r_{2}}^{\infty}dr_{1}r_{1}\sum_{l=0}^{\infty}\left(\frac{r_{2}}{r_{1}}\right)^{l}f(r_{1})\int P_{l}(\cos\theta)d\Omega\nonumber \\
 & = & \frac{4\pi}{r_{2}}\int_{0}^{r_{2}}dr_{1}r_{1}^{2}f(r_{1})+4\pi\int_{r_{2}}^{\infty}dr_{1}r_{1}f(r_{1})\label{eq:4}
\end{eqnarray}
where we have used
\begin{equation}
\int P_{l}(\cos\theta)d\Omega=2\pi\int_{-1}^{1}P_{l}(x)dx=2\pi\int_{-1}^{1}P_{l}(x)P_{0}(x)dx=4\pi\delta_{l0}\label{eq:5}
\end{equation}

After these preliminary work, the four integrals in Eq. (\ref{eq:A1})
can be calculated as follows:
\begin{eqnarray}
\mathrm{I}_{1} & = & \int dV_{2}\psi_{100}(r_{2})^{2}\int dV_{1}\frac{1}{r_{12}}\psi_{100}(r_{1})^{2}\nonumber \\
 & = & \int dV_{2}\frac{\lambda^{3}}{\pi a_{B}^{3}}e^{-2\lambda r_{2}/a_{B}}\frac{\lambda^{3}}{b^{2}a_{B}^{3}}\times\frac{1}{br_{2}}\left[1-(1+br_{2})e^{-2br_{2}}\right]\nonumber \\
 & = & \frac{5}{8}\times\frac{\lambda}{a_{B}}\label{eq:d1}
\end{eqnarray}
\begin{eqnarray}
\mathrm{I}_{2} & = & \int dV_{2}\psi_{200}(r_{2})^{2}\int dV_{1}\frac{1}{r_{12}}\psi_{100}(r_{1})^{2}\nonumber \\
 & = & \frac{\lambda^{3}}{\pi a_{B}^{\prime3}}\frac{\lambda^{3}}{b^{3}a_{B}^{3}}\int dV_{2}\left(1-\frac{br_{2}}{2}\right)^{2}e^{-br_{2}}\frac{1}{r_{2}}\left[1-(1+br_{2})e^{-2br_{2}}\right]\nonumber \\
 & = & \frac{17}{81}\times\frac{\lambda}{a_{B}}\label{eq:d2}
\end{eqnarray}
\begin{eqnarray}
\mathrm{I}_{3} & = & \int dV_{2}\psi_{100}(r_{2})\psi_{200}(r_{2})\int dV_{1}\frac{1}{r_{12}}\psi_{100}(r_{1})\psi_{200}(r_{1})\nonumber \\
 & = & \frac{\lambda^{3}}{\pi a_{B}^{3}}\frac{\lambda^{3}}{\pi a_{B}^{\prime3}}\frac{4\pi}{b^{2}}\frac{1}{27}\int dV_{2}\times\left(1-\frac{br_{2}}{2}\right)\left(4+6br_{2}\right)e^{-3br_{2}}\nonumber \\
 & = & \frac{16}{27^{2}}\times\frac{\lambda}{a_{B}}\label{eq:d3}
\end{eqnarray}
\begin{eqnarray}
\mathrm{I}_{4} & = & \int dV_{2}\psi_{200}(r_{2})^{2}\int dV_{1}\frac{1}{r_{12}}\psi_{200}(r_{1})^{2}\nonumber \\
 & = & \frac{\lambda^{3}}{\pi a_{B}^{\prime3}}\frac{\lambda^{3}}{b^{2}a_{B}^{\prime3}}\int dV_{2}\left(1-\frac{\lambda r_{2}}{a_{B}^{\prime}}\right)^{2}e^{-2\lambda r_{2}/a_{B}^{\prime}}\times\frac{1}{br_{2}}\left[8-\left(8+6br_{2}+2b^{2}r_{2}^{2}+b^{3}r_{2}^{3}\right)e^{-br_{2}}\right]\nonumber \\
 & = & \frac{77}{512}\times\frac{\lambda}{a_{B}}\label{eq:d4}
\end{eqnarray}

\section{The complete form of $\Psi^{\dagger}\Psi$ in Sce. \ref{sec:Be}}

\label{sec:psipsi} We list the complete form of $\Psi^{\dagger}\Psi$
in Sce. \ref{sec:Be} in the following:
\begin{eqnarray*}
\Psi^{\dagger}\Psi\times4! & = & \ \ (1234)^{2}+1234(-3214-1432+3412)\\
 &  & +(2314)^{2}+2314(-1324-2413+1423)\\
 &  & +(3124)^{2}+3124(-2134-3421+2431)\\
 &  & +(2134)^{2}+2134(-3124-2431+3421)\\
 &  & +(3214)^{2}+3214(-1234-3412+1432)\\
 &  & +(1324)^{2}+1324(-2314-1423+2413)\\
\\
 &  & +(1243)^{2}+1243(-4213-1342+4312)\\
 &  & +(2341)^{2}+2341(-4321-2143+4123)\\
 &  & +(3142)^{2}+3142(-4132-3241+4231)\\
 &  & +(2143)^{2}+2143(-4123-2341+4321)\\
 &  & +(3241)^{2}+3241(-4231-3142+4132)\\
 &  & +(1342)^{2}+1342(-4312-1243+4213)\\
\\
 &  & +(1423)^{2}+1423(-2413-1324+2314)\\
 &  & +(2431)^{2}+2431(-3421-2134+3124)\\
 &  & +(3412)^{2}+3412(-1432-3214+1234)\\
 &  & +(2413)^{2}+2413(-1423-2314+1324)\\
 &  & +(3421)^{2}+3421(-2431-3124+2134)\\
 &  & +(1432)^{2}+1432(-3412-1234+3214)\\
\\
 &  & +(4123)^{2}+4123(-2143-4321+2341)\\
 &  & +(4231)^{2}+4231(-3241-4132+3142)\\
 &  & +(4312)^{2}+4312(-1342-4213+1243)\\
 &  & +(4312)^{2}+4312(-1342-4213+1243)\\
 &  & +(4321)^{2}+4321(-2341-4123+2143)\\
 &  & +(4132)^{2}+4132(-3142-4231+3241)
\end{eqnarray*}

\bibliographystyle{apsrev}
\addcontentsline{toc}{section}{\refname}\bibliography{ref-20250507}

\end{document}